\documentclass[twocolumn,english,aps,pra,superscriptaddress]{revtex4-2}
\usepackage[utf8]{inputenc}
\usepackage{xcolor}
\usepackage{amsmath}
\usepackage{amssymb}
\usepackage{graphicx}

\makeatletter
\usepackage{babel}

\makeatother

\usepackage{babel}
\begin{document}
\title{Note on the noise reduction in spectroscopic detection with compressed
sensing}
\author{Junyan Sun}
\address{Graduate School of China Academy of Engineering Physics, Beijing 100193,
China}
\address{Beijing Computational Science Research Center, Beijing 100193, China}
\author{Deran Zhang}
\address{Graduate School of China Academy of Engineering Physics, Beijing 100193,
China}
\author{Ziqian Cheng}
\address{Graduate School of China Academy of Engineering Physics, Beijing 100193,
China}
\author{Dazhi Xu}
\email{dzxu@bit.edu.cn}

\address{Center for Quantum Technology Research and Key Laboratory of Advanced
Optoelectronic Quantum Architecture and Measurements (MOE), School
of Physics, Beijing Institute of Technology, Beijing 100081, China}
\author{Hui Dong}
\email{hdong@gscaep.ac.cn}

\address{Graduate School of China Academy of Engineering Physics, Beijing 100193,
China}
\begin{abstract}
Spectroscopy sampling along delay time is typically performed with
uniform delay spacing, which has to be low enough to satisfy the Nyquist--Shannon
sampling theorem. The sampling theorem puts the lower bound for the
sampling rate to ensure accurate resolution of the spectral features.
However, this bound can be relaxed by leveraging prior knowledge of
the signals, such as sparsity. Compressed sensing, a under-sampling
technique successfully applied to spatial measurements (e.g., single-pixel
imaging), has yet to be fully explored for the spectral measurements
especially for the temporal sampling. In this work, we investigate
the capability of compressed sensing for improving the temporal spectroscopic
measurements to mitigate both measurement noise and intrinsic noise.
By applying compressed sensing to single-shot pump-probe data, we
demonstrate its effectiveness in noise reduction. Additionally, we
propose a feasible experimental scheme using a digital mirror device
to implement compressed sensing for temporal sampling. This approach
provides a promising method for spectroscopy to reduce the signal
noise and the number of sample measurements.
\end{abstract}
\maketitle

\section{introduction}

In the spectroscopic measurement, the time-dependence of signal intensity
is typically obtained by a sequence of measurements with equal-spaced
time delays. For example, the signals of the transient absorption
spectroscopy are collected along the delay time between the pump and
the probe pulses with equal spacing. The minimum sampling rate for
accurate resolving the signal, as determined by the Nyquist--Shannon
sampling theorem \cite{Nyquist1928,Shannon1949}, is at least twice
the signal band width. To avoid subsampling, the sampling rate must
be set to a very high value based on the maximum oscillation frequency,
which is usually not precisely known. This requirement becomes even
more challenging in Fourier-transform-based spectral detection, such
as in frequency-domain measurements involving the first delay time
in two-dimensional spectroscopy \cite{Mukamel1995,Cho2009,Hamm2011,Jonas2003,SchlauCohen2011}.
Such measurements often require dense sampling in time to capture
rapid dynamics, resulting in long acquisition times and potential
challenges in experimental feasibility.

Compressed sensing \cite{Donoho2006,Duarte2011,Eldar2012,Kutyniok2013,Tsaig2006}
as an algorithmic assistant sampling method may offer a solution.
Considering most meaningful signals are sparse in certain representation
spaces, the equal-spaced Nyquist-Shannon sampling in such structured
spaces is often redundant. Compressed sensing provides an efficient
and generalized framework for sampling and recovering sparse signals,
offering the advantage of significantly reducing the number of measurements
\cite{Dunbar2013,Almeida2012} while remaining robust to noise \cite{Candes2006a}.
Based on these facts, compressed sensing has been rapidly applied
in magnetic resonance imaging \cite{Lustig2007}, nonlinear optical
imaging \cite{Cai2011}, multidimensional spectroscopy \cite{Sanders2012,Dunbar2013},
holography \cite{Rivenson2013} and super resolution microscopy \cite{Zhu2012},
and many more since it is proposed.

As one of the most prominent applications of compressed sensing in
spatially resolved signal acquisition, single-pixel camera enables
efficient imaging with far fewer measurements than traditional array-detector-based
methods \cite{Duarte2008}. This breakthrough has inspired a wide
range of novel imaging techniques \cite{Candes2006,Lakshminarayanan2014},
extending the applicability of compressed sensing beyond conventional
imaging systems \cite{Spencer2016}. Despite its success in spatial
sampling, the application of compressed sensing to temporal signal
acquisition is still in its early stages. A key challenge is developing
effective modulation techniques for broadband temporal signals. Some
efforts have been made, such as using driven atomic quantum systems
to modulate dynamic signals in quantum sensing \cite{Cappellaro2013,Xu2022}.
Additionally, the digital mirror device (DMD), widely used for spatial
signal modulation, has also been explored for compressive ultrafast
time-domain measurements \cite{Boyd2021}.

In this paper, we present a comprehensive analysis of applying compressed
sensing to the time-resolved spectroscopy measurements. While early
attempts have been made \cite{Adhikari2021}, the effectiveness of
compressed sensing in handling spectroscopic data with noises and
the feasibility of hardware implementation remains less explored.
Our study demonstrates that compressed sensing can effectively mitigate
the measurement noise originating from detection devices, but has
limited impact on reducing intrinsic noise encoded within the signal.
Using single-shot pump-probe transient absorption spectroscopy data
as a case study, we illustrate the capability of compressed sensing
in reducing signal noise. Furthermore, we propose a practical experimental
scheme for implementing compressive temporal sampling.

The remainder of this paper is organized as follows. Section II describes
the two-step processes for applying compressed sensing in spectroscopic
measurements. In Section III, we provide a detailed discussion of
the noise reduction effects for two types of noise. Section IV demonstrates
the application of compressed sensing to process single-shot experimental
data. In Section V, we present an experimental implementation of compressed
sensing in spectroscopic measurements. Finally, the main results are
summarized in Section VI.

\section{compressed sensing for spectroscopic measurements}

In the traditional spectroscopic measurements, the measured signal
typically is the desired signal itself or related to the desired signal
upon simple mathematical transformations, e.g., the Fourier transformation.
However, the compressed sensing offers a complete different strategy
with two-step processes, \textbf{data acquisition} and \textbf{signal
recovery}. In the data acquisition process, the data is collected
with designed linear modulation on the original signal. The dimension
of the acquired data is significantly reduced comparing to the dimension
determined by the the Nyquist--Shannon sampling theorem. In the signal
recovery process, the high dimensional signal is obtained from the
acquired low dimensional data by an optimization algorithm. The effectiveness
of the compressed sensing is ensured with the sparsity nature of the
signal, along with the sampling and recovery algorithm. We anticipate
that the application of the compressed sensing, along with many other
algorithms, will enable a new research realm for the spectroscopy,
i.e., the algorithmic spectroscopy.

In this section, we will introduce the two-step processes for the
compressed sensing, with the emphasis on the application in the temporal
sampling of the spectroscopy signal. To avoid any divergence of the
discussion, we will skip several aspects related to the mathematical
strictness of the compressed sensing as well as its generality.

We denote $f$ as the signal of interest with the dimension $N$,
i.e., $f\in\mathbb{R}^{N}$ as a vector. Such dimension should be
large enough to fulfill the Nyquist--Shannon sampling theorem. If
we consider the total signal duration as $T$, the sampling frequency
is $2\pi(N-1)/T$, which should be at least twice of the maximum frequency
of the signal. The compressed sensing consists the two steps as follows.

Step 1,\textbf{ data acquisition.} The data acquisition is carried
out via linear modulation of the signal $f$ through a sampling matrix
$\boldsymbol{A}$ as 
\begin{equation}
Y=\boldsymbol{A}f,\label{eq:1}
\end{equation}
where $\boldsymbol{A}\in\mathbb{R^{\mathit{K\times N}}}$ is the sampling
matrix with $K\ll N$. And $\mathfrak{\mathit{Y}}\in\mathbb{R}^{K}$
is the data vector acquired from the experimental measurements. In
the following discussion, we assume that the signal in the spectroscopy
experiments is of the form 
\begin{equation}
f(t)=e^{-\gamma t}(\cos\omega t+a),\label{eq:SignalForm}
\end{equation}
where $\gamma$ is damping rate, $\omega$ is frequency of the signal,
and $a$ is non-zero constant shift. Traditionally, the signal vector
$f$ is obtained with sampling at time location\textcolor{teal}{{} }$t_{i}=t_{1}+(i-1)T/(N-1)$
as $f_{i}=f(t_{i})$ with $i=1,...,N$.

Typically, the sampling matrix $\boldsymbol{A}$ should be generated
to fulfill rigorous mathematical conditions such as the restrict isometric
properties (RIP) \cite{Candes2005}. Fortunately, it has been proved
a randomly generated matrix, with distribution such as Gaussian \cite{Chen2005},
Bernoulli \cite{Candes2006}, etc., has very high probability to satisfy
the RIP \cite{Donoho2006}. In this study, we adopt the simple binary
random Bernoulli sensing matrix, whose element $\boldsymbol{A}_{ij}$
is $0$ or $1$ with the probability $p$ or $1-p$, $p\in(0,1)$.
Under this special choice of the sampling matrix, the acquired data
$Y$ of dimension $K$ is the sum of the signal at the random time
points. In the case with generic random matrix $\boldsymbol{A}$,
the obtained data $Y$ is a linear combination of all the time points.
Such acquired data is different from the data collected with the traditional
sampling methods, where the major efforts are devoted to resolve different
time points rather than to combine them together. With the acquired
data, the next task is to recover the desired signal. Mathematically,
this task can be described as exactly recovering a $N$ dimensional
vector from a $K$ ($K\ll M$) dimensional vector $Y$ obtained by
a low-rank linear transformation.

Step 2, \textbf{signal recovery}. It seems there are infinite possible
reconstructed signal $\hat{f}$ considering the linear equations $\boldsymbol{A}f=Y$
are under-determined. However, the compressed sensing theory takes
advantage of the sparsity of the natural signal. The signal $f$ of
interest can be transformed into a representation by $f^{\prime}=\Phi f$,
where $\Phi$ is a transformation matrix with dimension $N\times N$
and $f^{\prime}$ is a sparse vector. With such transformation, the
collected data is rewritten as 
\begin{equation}
Y=\boldsymbol{A}\Phi^{-1}\Phi f=\boldsymbol{B}f^{\prime},\label{eq:recover}
\end{equation}
where $\boldsymbol{B}=\boldsymbol{A}\Phi^{-1}$ is the reconstruction
matrix with dimension $K\times N$.

The signal $\hat{f}^{\prime}$ in the sparse space is reconstructed
by minimizing the $l_{1}-$norm $\hat{f}^{\prime}=\arg_{f^{\prime}}\min\left\Vert f^{\prime}\right\Vert _{1}$
subject to $\boldsymbol{B}f^{\prime}=Y$, where $\left\Vert f^{\prime}\right\Vert _{1}=\sum_{i=1}^{N}\left|f_{i}^{\prime}\right|$.
The recovered signal is finally obtained by $\hat{f}=\Phi^{-1}\hat{f}^{\prime}$.
The fidelity between the recovered signal and the actual signal relies
significantly on the choice of the transformation matrix $\Phi$.
It has been shown that the discrete cosine transformation (DCT), Hadamard
transformation, and fast Fourier transformation are effective in the
signal recovery \cite{Qaisar2013}. In the current paper, we find
the DCT, whose matrix element reads 
\begin{equation}
\Phi_{ij}=\cos[\frac{\pi}{N}(i-\frac{1}{2})(j-1)],\label{eq:dct}
\end{equation}
is effective for signal recovery in the spectroscopic measurements.

\section{The effect of compressed sensing on the noise}

Considering the real experimental data are all noisy, the compressed
measurement samples the weighted average of the original signal together
with the noise. Because the noise has been averaged during the sampling
procedure, it can be expected that the compressed sensing can improve
the signal to noise ratio. In the experimental measurements, the noise
is typically divided into the two categories, the intrinsic noise
associated with the signal and the noise associated with the measurement
devices. Accordingly, we analysis the noise reduction effects of compressed
sensing in this section. We rewrite the acquired spectroscopic data
as 
\begin{equation}
Y=\boldsymbol{A}(f+\delta_{1})+\delta_{2},
\end{equation}
where $\delta_{1}\in\mathbb{R}^{N}$ calibrates the intrinsic noise
of the light field to be measured, and $\delta_{2}\in\mathbb{R}^{K}$
represents the noise generated by the measurement devices. \textcolor{black}{The
intrinsic noise $\delta_{1}$ is assumed independent of the measurement
noise $\delta_{2}$. The }intrinsic noise $\delta_{1}$\textcolor{black}{{}
may result from the unstable light source, the dynamic changes of
the transmission optical path (e.g., the air flow caused by choppers
in the pump-probe experiment), the inherent instability of the samples,
and the scattering of the pump light on the sample surface. The measurement
noise $\delta_{2}$ refers to the noise from the measurement system,
mainly from the inherent electrical noise of the instrument itself
and the scattering light (e.g. pump and environmental background light)
captured by the apparatus. We remark that the scattering light in
the noise $\delta_{2}$ is dispersed throughout the array detector,
which is different from the noise $\delta_{1}$ encoded by the the
sampling matrix $\boldsymbol{A}$. }Here, we investigate the two types
of noise (i.e., $\delta_{1}$ and $\delta_{2}$) in the signal obtained
by the compressed sensing with numerical simulations. For comparison,
we also define the Nyquist-Shannon sampled noisy signal as $f_{\mathrm{NS}}=f+\delta_{1}+\delta_{2}$.

\subsection{The impact on the measurement noise $\delta_{2}$}\label{subsec:impact_delta2}

We firstly consider the impact on the measurement noise $\delta_{2}$
by setting $\delta_{1}=0$. The data sampled is written as 
\begin{equation}
Y=\boldsymbol{A}f+\delta_{2}.\label{eq:delta2}
\end{equation}
In the simulation, we have used the signal in Eq. (\ref{eq:SignalForm})
with the decay rate $\gamma=20\textrm{GHz}$ corresponding to lifetime
$50\textrm{ps}$, the oscillation frequency $\omega=3.77\times10^{15}\textrm{rad/s}$
corresponding to wavelength $500\textrm{nm}$ and the shift $a=5$.
It is assumed that the measurement noise $\delta_{2}$ follows a Gaussian
distribution with mean of zero and a variance of $\sigma$. The total
signal dimension is $N=1024$ and the compressed sensing sampling
dimension is $K=200$. With the simulated data $Y$, we reconstruct
the signal $\hat{f}$ with the orthogonal matching pursuit algorithm
\cite{Dai2009,Qaisar2013,Eldar2012}. We use the root-mean-square
error 
\begin{equation}
\mathrm{RMSE}(\hat{f},f)=\sqrt{\frac{1}{N}\sum_{i=1}^{N}|\hat{f}_{i}-f_{i}|^{2}}
\end{equation}
to evaluate the fidelity between the reconstructed signal $\hat{f}$
and the original signal $f$. The Nyquist-Shannon sampled noisy signal
is $f_{\mathrm{NS}}=f+\delta_{2}$, whose root-mean-square error is
$\mathrm{RMSE}(f_{\mathrm{NS}},f)$.

\begin{figure}
\includegraphics{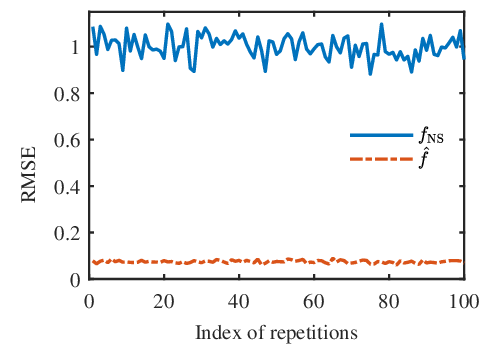}\caption{The root-mean-square error of Nyquist--Shannon sampled signal $f_{\mathrm{NS}}$
(blue solid line) and the the reconstructed signal $\hat{f}$ (red
dash-dotted line). The measurement noise $\delta_{2}$ follows a Gaussian
distribution with mean zero and variance $\sigma=1$. The parameters
are $\gamma=20\textrm{GHz}$, $\omega=3.77\times10^{15}\textrm{rad/s}$,
$a=5$, $N=1024$, $K=200$.}

\label{fig:delta2_rmse_repetition}
\end{figure}

Fig. \ref{fig:delta2_rmse_repetition} shows the the root-mean-square
error of the recovered signal $\hat{f}$ with compressed sensing (red
dash-dotted line) and the signal $f_{\mathrm{NS}}$ from Nyquist-Shannon
sampling (blue solid line) as functions of the index of different
repetitions. The variance of the noise is chosen as $\sigma=1$. For
the Nyquist-Shannon sampling, the standard deviation is approximate
$\sigma$ with fluctuations. And the noise in the compressive sampled
signal is significantly reduced by one order of magnitude, which implies
that compressed sensing has potential in reducing the noise. The noise
was proved to be effectively controlled to allow the reliable recovery
\cite{Candes2006a}. Here, we show that in the spectroscopy measurement,
such noise can be further reduced by the choosing the proper sampling
matrix $\boldsymbol{A}$ and the transformation matrix $\Phi$.

\begin{figure}
\includegraphics{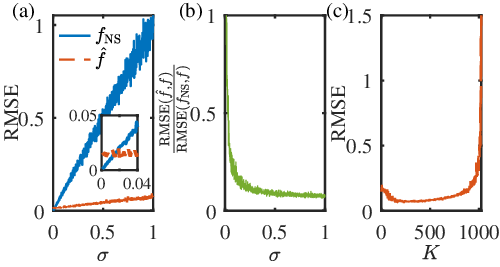}\caption{The root-mean-square error of the Nyquist--Shannon sampled signal
$f_{\mathrm{NS}}$ (blue solid line) and the reconstructed signal
$\hat{f}$ (red dashed line) as functions of (a) the variance $\sigma$
of $\delta_{2}$ and (c) the sampling dimension $K$. (b) shows the
ratio of $\mathrm{RMSE}(\hat{f},f)$ and $\mathrm{RMSE}(f_{\mathrm{NS}},f)$
as a function of the variance $\sigma$ of $\delta_{2}$. The parameters
chosen are the same as Fig. \ref{fig:delta2_rmse_repetition}. }\label{fig:delta2_rmse_noiselevel}
\end{figure}

To qualitatively evaluate the impact on the measurement noise, we
simulate the noise reduction for different variance $\sigma$ in Figs.
\ref{fig:delta2_rmse_noiselevel}(a,b) and the sampling dimension
$K$ in Fig. \ref{fig:delta2_rmse_noiselevel}(c). In Fig. \ref{fig:delta2_rmse_noiselevel}(a),
we plot the root-mean-square error of the recovered signal $\hat{f}$
(red dashed line) and the Nyquist-Shannon sampled signal $f_{\mathrm{NS}}$
(blue solid line) as a function of the variance $\sigma$. For the
weak noise ($\sigma<0.02$), the compressed sensing shows no effect
on reducing the noise, since the fluctuation introduced by the sampling
matrix which is randomly regenerated for each signal recovery. However,
for most cases, the application of compressed sensing significantly
reduces the noise by one order of magnitude, as illustrated in Fig.
\ref{fig:delta2_rmse_noiselevel}(b), where the ratio of the noise
reduction is plotted as a function of the variance $\sigma$. Therefore,
compressed sensing is superior to Nyquist-Shannon sampling in terms
of reducing measurement noise.

In Fig. \ref{fig:delta2_rmse_noiselevel}(c), we plot the root-mean-square
error of the recovered signal $\hat{f}$ versus the sampling dimension
$K$. For the small sampling dimension $K$, the dimension of the
acquired data $Y$ is not large enough to provide information in the
sparse domain to reconstructed the signal. For the large sampling
dimension $K$, the recovery algorithm in general will recover the
signal as well as the noise, and results in the large noise. The theory
of compressed sensing has predicted the minimum sampling dimension
$K\sim O(s\log N)$ \cite{Dai2009,Qaisar2013,Chen2001}, where $s$
is the non-zero number of the sparse signal $f^{\prime}$ (i.e., $f^{\prime}$
is $s$-sparse) in Eq. (\ref{eq:recover}). The minimum sampling dimension
$K$ is required for basis pursuit algorithm based on linear programming,
and applying different reconstruction algorithm will result in different
minimum sampling requirement \cite{Qaisar2013}.

\subsection{The impact on the intrinsic noise $\delta_{1}$}\label{subsec:impact_delta1}

Now we turn to the impact on intrinsic noise $\delta_{1}$ as 
\begin{equation}
Y=\boldsymbol{A}(f+\delta_{1}).\label{eq:delta1}
\end{equation}
Similarly, we set the intrinsic noise $\delta_{1}$ as a zero-mean
Gaussian random vector with the variance $\sigma$, and assume the
signal collected by the Nyquist-Shannon sampling method as $f_{\mathrm{NS}}=f+\delta_{1}$
with $\delta_{2}=0$.

\begin{figure}
\includegraphics{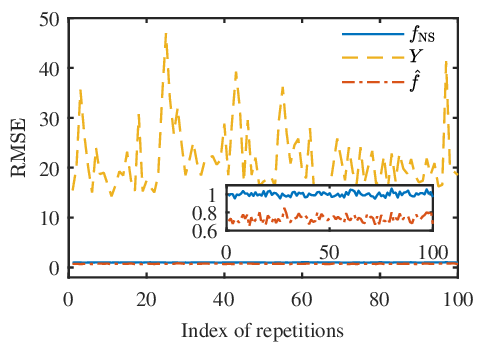}

\caption{The root-mean-square error of the Nyquist--Shannon sampled signal
$f_{\mathrm{NS}}$ (blue solid line), the reconstructed signal $\hat{f}$
(red dash-dotted line), and the collected signal $Y$ (yellow dashed
line). The root-mean-square error of collected data $Y$ is defined
as $\mathrm{RMSE}(Y,\boldsymbol{A}f)$. The intrinsic noise $\delta_{1}$
follows a Gaussian distribution with mean zero and variance $\sigma=1$.
The other parameters chosen are the same as Fig. \ref{fig:delta2_rmse_repetition}.
}\label{fig:delta1_rmse_repetition}
\end{figure}

Fig. \ref{fig:delta1_rmse_repetition} shows the root-mean-square
error of the recovered signal $\hat{f}$ with compressed sensing (red
dot-dashed line), and the signal $f_{\mathrm{NS}}$ from the Nyquist-Shannon
sampling method (blue solid line). The standard deviation of the signal
from the Nyquist-Shannon sampling with intrinsic noise is approximate
$\sigma$, while the signal recovered from compressed sensing is reduced
about $30\%$. It is clear that the compressed sensing has limited
capability on reducing the intrinsic noise, compared to the measurement
noise $\delta_{2}$. To explain this observation, we plot the root-mean-square
error $\mathrm{RMSE}(Y,\boldsymbol{A}f)$ of the acquired data $Y$
in Fig. \ref{fig:delta1_rmse_repetition}. The noise in one particular
acquired data $Y_{i}$ is large due to the combination of noise from
different time points $\sum_{j}\boldsymbol{A}_{ij}(\delta_{1})_{j}$,
which mitigates the reduction effect on intrinsic noise.

\begin{figure}
\centering \includegraphics{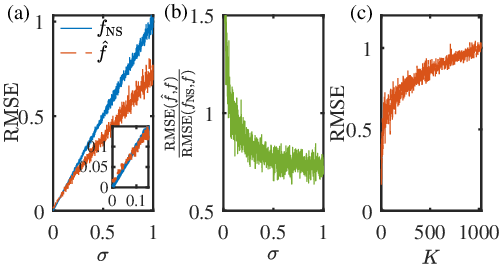}\caption{The root-mean-square error of the Nyquist-Shannon sampled signal $f_{\mathrm{NS}}$
(blue solid line) and the reconstructed signal $\hat{f}$ (red dashed
line) as functions of (a) the variance $\sigma$ of $\delta_{1}$
and (c) the sampling dimension $K$. (b) shows the ratio of $\mathrm{RMSE}(\hat{f},f)$
and $\mathrm{RMSE}(f_{\mathrm{NS}},f)$ as a function of the variance
$\sigma$ of $\delta_{1}$. The parameters chosen are the same as
Fig. \ref{fig:delta1_rmse_repetition}. }\label{fig:delta1_rmse_noiselevel_k}
\end{figure}

To quantitatively evaluate the impact of compressed sensing, we plot
the root-mean-square error of the recovered signal and the signal
sampled by the Nyquist-Shannon sampling as functions of the noise
variance $\sigma$ of $\delta_{1}$ in Figs. \ref{fig:delta1_rmse_noiselevel_k}(a,b)
and as functions of the sampling dimension $K$ in Fig. \ref{fig:delta1_rmse_noiselevel_k}(c).
Similar to that of the measurement noise, the reduction of compressed
sensing on large intrinsic noise is prominent with a $30\%$ reduction
compared to the Nyquist-Shannon sampling as shown in Figs. \ref{fig:delta1_rmse_noiselevel_k}(a)
and (b). Fig. \ref{fig:delta1_rmse_noiselevel_k}(c) illustrates that
large sampling dimension $K$ leads to a decrease of the noise reduction.

In this section, we have demonstrated the noise reduction in the compressed
sensing for two types of the noise, i.e., the measurement noise $\delta_{2}$
and the intrinsic noise $\delta_{1}$. The compressed sensing shows
the significant reduction of the measurement noise $\delta_{2}$ by
one order of magnitude, while has the mild reduction of the intrinsic
noise $\delta_{1}$ around $30\%$. Indeed, the two types of the noise
can be treated equally with the so-called noise folding \cite{AriasCastro2011},
which has been mathematically investigated. We have intentionally
avoid such mathematical discussion in the current paper to focus on
the spectroscopic applications.

We note that the spectroscopic signals are assumed to be sparse in
the DCT domain as defined in Eq. (\ref{eq:dct}). This sparse representation
is selected based on a comparison of recovery results with those obtained
from various other transformations. Additionally, we demonstrate the
existence of an optimal sampling dimension \ensuremath{K} that minimizes
measurement noise in the spectroscopy application.

\section{Application of compressed sensing in the single-shot experiment}

In the last section, we have numerically verified the noise reduction
effects of the compressed sensing. Based on this fact, in this section,
we will use the compressed sensing as a data process method to reconstruct
the signal acquired from the Nyquist-Shannon sampled single-shot experimental
data.

\begin{figure}
\begin{centering}
\includegraphics{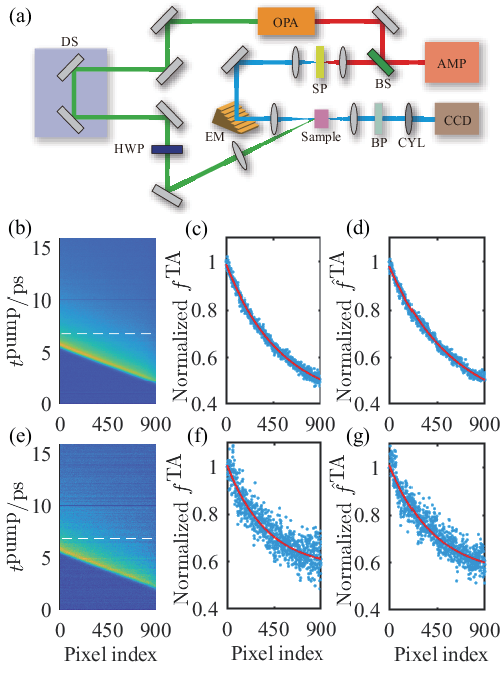}
\par\end{centering}
\caption{(a) Schematic of the single-shot pump\textcolor{black}{\protect\nobreakdash-}probe
TA experiment. AMP: Ti:Sapphire amplifier laser system; BS: beam splitter;
OPA: optical parametric amplifier; DS: delay stage; HWP: half-wave
plate; SP: sapphire window; EM: echelon mirror; BP: bandpass filter;
CYL: cylindrical lens; CCD: charge-coupled device camera. In the entire
diagram, the red line, blue line, and green line represent the $800\textrm{nm}$
fundamental pulse, the probe pulse, and the pump pulse, respectively.
The single-shot experimental data $f^{\text{TA}}$ acquired from Nyquist-Shannon
sampled method with averaging over (b) 1000 and (e) 100 pulses. For
averaging 1000 pulses, (c) and (d) show the normalized single-shot
TA signal $f^{\text{TA}}$ and the normalized recovered signal $\hat{f}^{\text{TA}}$
at time delay $t^{\textrm{pump}}=6.9\mathrm{ps}$ denoted by white
dashed line in (b). (f) and (g) correspond to normalized $f^{\text{TA}}$
and normalized $\hat{f}^{\text{TA}}$ for averaging 100 pulses at
$t^{\textrm{pump}}=6.9\mathrm{ps}$ (white dashed line in (e)). The
data in the figure are normalized by the intensity at the first pixel
and when pump and probe pulses completely overlap. The red solid curves
denote fits to the acquired signal $f^{\text{TA}}$ or the recovered
signal $\hat{f}^{\text{TA}}$ of the same form $f_{\textrm{fit}}^{\text{TA}}(t)=\hat{f}_{\textrm{fit}}^{\text{TA}}(t)=c_{1}e^{-t/\tau}+c_{2}$
with different fitting parameters (c) $c_{1}=0.5891,c_{2}=0.4013,\tau=523.6$;
(d) $c_{1}=0.6159,c_{2}=0.3659,\tau=606.5$; (f) $c_{1}=0.4375,c_{2}=0.5704,\tau=376.2$;
(g) $c_{1}=0.4659,c_{2}=0.5463,\tau=417.7$.}\label{fig:data}
\end{figure}

Fig. \ref{fig:data}(a) illustrates the schematic of our single-shot
pump\nobreakdash-probe transient absorption (TA) experiment. The
output light from the source is split into two beams: one is directed
through an optical parametric amplifier (OPA) to generate the pump
pulse, while the other passes through a sapphire window to produce
supercontinuum white light, which serves as the probe pulse. The acquired
data in the single-shot TA experiment is expressed as the relative
transmitted ratio, $f^{\text{TA}}=\Delta I_{\text{T}}/I_{\text{T}}$,
where $I_{\text{T}}$ is the transmitted intensity of the probe pulse
in absence of the pump pulse, and $\Delta I_{\text{T}}$ represents
the transmitted intensity difference between the probe pulse with
and without the pump pulse.

The pump pulse is passed through an electronically controlled mechanical
delay stage to adjust its delay time $t^{\textrm{pump}}$. The total
delay range of the stage is $16\textrm{ps}$, with a time step of
$\delta t^{\textrm{pump}}=26.7\textrm{fs}$. The probe pulse is incident
on an echelon mirror and reflected as a sequence of sub-pulses. The
step size of the echelon mirror is $7.5\mu\mathrm{m}$, resulting
in a delay of $\delta t^{\textrm{prob}}=50\mathrm{fs}$ between consecutive
sub-pulses. The total delay range covered by these sub-pulses is $3.5\mathrm{ps}$.
These sub-pulses, each corresponding to a different delay time $t^{\textrm{prob}}$,
propagate through the sample and a cylindrical lens before being detected
by a one-dimensional charge-coupled device (CCD) camera with $N=901$
pixels. Consequently, the TA signal spanning a $3.5\mathrm{ps}$ delay
window is generated by a single probe pulse, with the delay time information
encoded in the spatial positions of the CCD pixels.

It should be note that the each step of the echelon mirror and the
CCD pixel do not correspond to a one-to-one mapping. The delay between
consecutive sub-pulses slightly less than $50\mathrm{fs}$ actually
since the probe pulse is not vertically incident on the echelon mirror.
Additionally, interference between adjacent steps of the echelon mirror
further complicates the calibration of the delay time $t^{\textrm{prob}}$
for each signal recorded by the CCD. Therefore, we denote the experimental
signal with pump pulse delay $t^{\textrm{pump}}$, recorded at the
$j-\textrm{th}$ pixel of the CCD, as $f_{j}^{\text{TA}}(t^{\textrm{pump}})$,
without explicitly indicating its dependence on $t^{\textrm{prob}}$.

We also emphasize that different combinations of delay times $t^{\textrm{pump}}$
and $t^{\textrm{prob}}$ can correspond to the same total delay time
after the pump pulse, leading to redundancy in the experimental data.
In the present experiment, this redundant data is utilized for both
calibrating the actual delay time $t^{\textrm{prob}}$ and verifying
the reliability of the single-shot signal.

Since the acquired signal $f^{\text{TA}}$ inevitably includes noise
along with the desired signal, we average over multiple pulses to
reduce the noise. The single-shot TA data $f^{\text{TA}}$ are presented
as functions of the delay time $t^{\textrm{pump}}$ and the CCD pixel
index $j\in[1,N]$ in Figs. \ref{fig:data}(b) and (e). These plots
are obtained by averaging over 1000 and 100 pulses for each fixed
value of $t^{\textrm{pump}}$, respectively. As expected, averaging
over a larger number of pulses results in lower relative signal noise.
This is clearly demonstrated in Figs. \ref{fig:data}(c) and (f),
where the single-shot data are extracted from Figs. \ref{fig:data}(b)
and (e), respectively, at delay time $t^{\textrm{pump}}=6.9\mathrm{ps}$
(denoted by the white dashed lines). The data in the figure are normalized
by the intensity at the first pixel and when pump and probe pulses
completely overlap. The red solid curves in Figs. \ref{fig:data}(c)
and (f) represent fits to the acquired signal $f^{\text{TA}}$ of
the form $f_{\textrm{fit}}^{\text{TA}}(t)=c_{1}e^{-t/\tau}+c_{2}$,
where $c_{1},$ $c_{2}$ and $\tau$ are fitting parameters. The root-mean-square
error between the acquired signal and the fit, $\mathrm{RMSE}(f^{\text{TA}},f_{\textrm{fit}}^{\text{TA}})$,
is $0.0140$ for the data averaged over 1000 pulses and $0.0471$
for the data averaged over 100 pulses.

In the single-shot experiment, our goal is to reduce the number of
the averaged pulses in order to save acquisition time, while still
maintaining a relatively low level of signal noise. To achieve this,
we employ compressed sensing to process the single-shot TA experimental
data. For each fixed pump delay $t^{\textrm{pump}}$, the single-shot
TA data are discretized into a vector with $N=901$ elements. The
sampling matrix $\boldsymbol{A}$ in Eq. (\ref{eq:1}) is designed
as a uniform random binary matrix of size $K\times N$, with $K=150$.
Each Nyquist-Shannon sampled signal $f^{\text{TA}}$ is a compressed
to a lower-dimensional vector $Y=\boldsymbol{A}f^{\text{TA}}$. Using
$Y$, $\boldsymbol{A}$ and the DCT matrix in Eq. (\ref{eq:dct}),
we reconstruct the signal $\hat{f}^{\text{TA}}$, which is expected
to exhibit a lower noise level compared to the original $f^{\text{TA}}$.

The reconstructed signals $\hat{f}^{\text{TA}}$ are shown in Figs.
\ref{fig:data}(d) and (g), corresponding to the Nyquist-Shannon sampled
signal $f^{\text{TA}}$ in (c) and (f). The red solid curves represent
the fitted vectors $\hat{f}_{\textrm{fit}}^{\textrm{TA}}$, which
follow the same function form as $f_{\textrm{fit}}^{\text{TA}}$,
but with different fitting parameters. The root-mean-square error
$\mathrm{RMSE}(\hat{f}^{\text{TA}},\hat{f}_{\textrm{fit}}^{\textrm{TA}})$
between the reconstructed signal $\hat{f}^{\text{TA}}$ and its fitted
counterpart $\hat{f}_{\textrm{fit}}^{\textrm{TA}}$ is $0.0136$ for
the data averaged over 1000 pulses, corresponding to a modest reduction
in noise of $2.9\%$ compared to the RMSE of $0.0140$ for the raw
$f^{\text{TA}}$. This noise reduction is consistent with the discussion
for the intrinsic noise in Sec. \ref{subsec:impact_delta1}. In contrast,
the $\mathrm{RMSE}(\hat{f}^{\text{TA}},\hat{f}_{\textrm{fit}}^{\textrm{TA}})$
for the data averaged over 100 pulses is $0.0408$, resulting in a
more significant noise reduction of $13.4\%$ compared to the RMSE
of $0.0471$ for the raw $f^{\text{TA}}$. This outcome suggests that
even with limited acquisition time, compressed sensing can effectively
mitigate noise to some extent.

However, it is important to highlight that the noise reduction in
the current single-shot experiment is relatively modest. This is primarily
because the noise is treated as intrinsic noise $\delta_{1}$, following
the Nyquist-Shannon sampling procedure. As shown in Sec. \ref{subsec:impact_delta1},
compressed sensing provides a valuable noise reduction tool, particularly
for measurement noise $\delta_{2}$, but its effectiveness in dealing
with intrinsic noise $\delta_{1}$ is limited. To fully exploit the
potential of compressed sensing for noise reduction, it is crucial
to integrate the compressed sensing technique into the signal acquisition
process itself, rather than applying it post-acquisition. In the following
section, we will describe how to implement compressed sensing experimentally
during data acquisition to achieve more substantial noise reduction
and optimize measurement time.

\begin{figure}
\centering \includegraphics{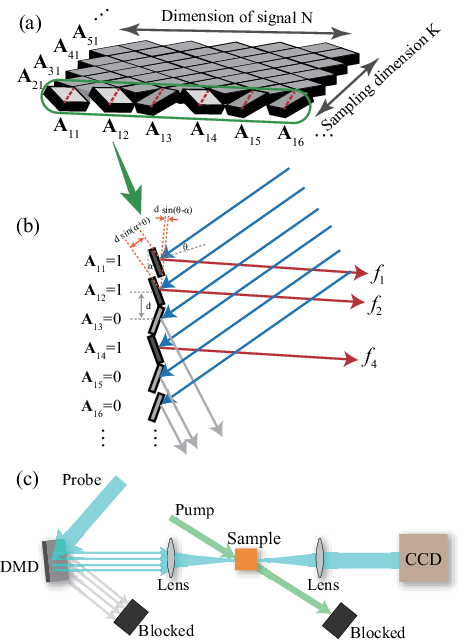}\caption{(a) Schematic diagram of DMD. (b) Schematic of one row of DMD in the
optical path. The blue line denotes the incident probe light. The
red line denotes the light as the probe beam reflected by the dark-gray
micromirrors toward the CCD, while the gray line corresponds to the
light reflected by the light-gray micromirrors away from the CCD.
The optical path difference of the incident beams between two adjacent
micromirrors is $\Delta d_{1}=d\sin(\alpha+\theta)$, and that of
the reflected beams is $\Delta d_{2}=d\sin(\theta-\alpha)$, resulting
in the total optical path difference $\Delta d=2d\sin\alpha\cos\theta$.
Here, $d$ is the distance between the center of two adjacent micromirrors.
$\theta$ is the incident angle of the light. $\alpha$ is the tilt
angle of the micromirrors. (c) Schematic of the experimental setup.
}\label{fig:DMD}
\end{figure}

\section{An experimental scheme for data acquisition}

In this section, we propose an experimental implementation of the
compressed sensing-based single-shot TA experiment, utilizing appropriate
hardware. The DMD, comprising millions of individually tiltable micromirrors
\cite{Sampsell1994}, will be employed to encode the sampling matrix
$\boldsymbol{A}$, as is typically done in imaging applications. As
depicted in Fig. \ref{fig:DMD}(a), each micromirror of the DMD can
tilt around an axis (red dashed line) to two fixed angles $\pm12^{\circ}$
(for most current DMDs), enabling the probe light to be reflected
into two distinct directions. To illustrate, consider one row of micromirrors,
as shown in Fig. \ref{fig:DMD}(b): the probe pulse incident on the
dark-gray micromirrors is directed toward the sample, ultimately reaching
the CCD for detection. In contrast, the micromirrors represented in
light-gray deflect the light away from the CCD, preventing detectioin.
Consequently, the DMD array functions as an optical mask for the received
signal, with each micromirror either reflecting the probe pulse toward
the CCD (adding a weight of 1 to the signal, for dark-gray micromirrors)
or away from the CCD (adding a weight of 0 to the signal, for light-gray
micromirrors).

In our experimental configuration, we use the $N$ micromirrors in
a single row of the DMD to generate a sequence of modulated sub-pulses
when a probe pulse is incident. Only those sub-pulses reflected by
the dark-gray micromirrors (with weight of 1) interact with the sample
and are detected by the CCD. Mathematically, this operations is equivalent
to one row of the matrix $\boldsymbol{A}$ acting on the signal. These
single-shot measurements are carried out simultaneously $K$ times
by utilizing $K$ rows of the micromirrors array. If $K\ll N$ and
the micromirrors are tilted randomly, the DMD effectively generates
a mask for the probe light, which can be described by a $K\times N$
dimensional Bernoulli random sampling matrix $\boldsymbol{A}$, where
each matrix element $\boldsymbol{A}_{ij}$ is either $0$ or $1$.

The tilting of the micromirrors also introduces an optical path difference
between the light reflected from adjacent micromirrors in a row. As
shown in Fig. \ref{fig:DMD}(b), this optical path difference is given
by $\Delta d=2d\sin\alpha\cos\theta$, where $d$ is the distance
between adjacent micromirrors, $\theta$ is the incident angle of
the probe light, and $\alpha$ is the tilt angle of the micromirrors.
Consequently, the spatial positions of the pixels on the CCD are correlated
with the probe light reflected from different micromirrors, which
allows the system to record the optical path differences or time delays
between the probe and pump pulses. We denote the signal corresponding
to the $j-\mathrm{th}$ micromirror as $f_{j}=f(t_{j})$, where $t_{j}$
represents the time delay induced by the tilt of the $j-\mathrm{th}$
micromirror.

For clarity, let us focus on the micromirrors in the first row of
the DMD. As depicted in Fig. \ref{fig:DMD}(c), when a probe pulse
illuminates this row, the modulated sub-pulses with weight of 1 are
focused by a lens and directed to the sample for pump-probe measurements.
The signals from different micromirrors are then directed to distinct
areas of the CCD for intensity measurement. The data we used for compressed
sensing signal recovery is the sum of the intensities on these pixels,
expressed as $Y_{1}=\sum_{j}\boldsymbol{A}_{1j}f_{j}$, where $Y_{1}$
is the measurement corresponding to the first row of the DMD. The
remaining data $Y_{i}$, for $i=2,\cdots,K$, are acquired from the
other rows of the DMD, resulting in the data vector $Y=[Y_{1},Y_{2},\cdots,Y_{K}]^{T}$
for further signal reconstruction algorithm.

In this setup, the intrinsic noise $\delta_{1}$ refers to the noise
encoded by the DMD in the spatial domain, which is reflected by the
micromirrors and captured by the CCD together with the signal $f$.
The measurement noise $\delta_{2}$, on the other hand, is independent
of DMD's coding process and includes contributions from the inherent
electrical noise of the measurement instrument and from the scattering
light. By employing the compressed sensing method outlined here, the
measurement noise can be significantly reduced, as demonstrated in
Sec. \ref{subsec:impact_delta2}.

\section{conclusion}

In summary, we have explored the role of compressed sensing in mitigating
noise in spectroscopy measurements involving temporal sampling. Our
findings demonstrate that while the compressed sensing is capable
of significantly reducing the measurement noise, it offers only moderate
improvements in mitigating the intrinsic noise. Through its application
to real single-shot pump\nobreakdash-probe transient absorption experimental
data, we show how compressed sensing can effectively suppress the
intrinsic noise in such measurements. Additionally, we present a practical
experimental implementation of the compressed sensing using a digital
micromirror device. Our work paves the way for the integration of
compressed sensing techniques into temporal sampling schemes for spectroscopic
experiments, offering potential advancements in data acquisition efficiency
and signal quality.
\begin{acknowledgments}
This work is supported by the Innovation Program for Quantum Science
and Technology (Grant No. 2023ZD0300700), and the National Natural
Science Foundation of China (Grant Nos. U2230203, U2330401, and 12088101).
\end{acknowledgments}

\bibliographystyle{apsrev4-2}
\bibliography{ref}

\end{document}